\documentstyle[preprint,epsfig,aps]{revtex}

\begin{document}

\draft

\tightenlines

\title{Quantum open systems and turbulence}

\author{L. F. Santos $^1$ and C. O. Escobar $^2$}

\address{$^1$ Sloane Physics Laboratory \\
Yale University \\
217 Prospect St, New Haven - CT 06520-8120, USA \\
santos@nst4.physics.yale.edu \\
$^2$ Departamento de  Raios C\'osmicos e Cronologia \\
Instituto de F\'{\i}sica Gleb Wataghin \\
Universidade Estadual de Campinas, C.P. 6165,  cep 13083-970\\
Campinas, S\~ao Paulo, Brazil\\
escobar@ifi.unicamp.br}

\maketitle

\begin{abstract}

We show that the problem of non conservation of energy found in
the spontaneous localization model developed by Ghirardi, Rimini
and Weber \cite{GRW} is very similar to the inconsistency between
the stochastic models for turbulence and the Navier-Stokes
equation. This sort of analogy may be useful in the development of
both areas.
\\
\end{abstract}


\section{Introduction}

Some models have been developed to describe how classical behavior
is regained from quantum mechanics, in an attempt to derive a
unified description of physical phenomena at any level. Quantum
mechanics allows the occurrence of linear superpositions of
macroscopically distinguishable states, which are not observed in
real life. A possible way to avoid them is by replacing the
unitary evolution of quantum mechanics by a non-unitary equation.

This was done for example by Ghirardi, Rimini and Weber (GRW)
\cite{GRW} and later improved by Ghirardi, Pearle and Rimini
\cite{GPR}. Here the Schr\"odinger equation is fundamentally
modified having now a stochastic term, which causes the collapse
of the wave function. Another possible modification of the von
Neumann evolution equation is obtained by considering the
interaction between the system of interest and a heat bath, as
done in the decoherence models \cite{Zurek,Teg,Gardiner}.

The master equation of the GRW model as well as the master
equation of some old versions of the decoherence models \cite{Zeh}
do not have a dissipation term, which leads to non conservation of
energy. Even though its is argued that this problem can be
disregarded in the GRW model since the energy production is very
small and probably unobservable \cite{GRW,Ballentine} and under
certain limits the dissipation term can in fact be neglected in
the equation for open systems \cite{Gallis}, we show here that the
increase of energy found in these models is very similar to the
inconsistency between the stochastic models for turbulence and the
Navier-Stokes equation.

Our discussion is done in terms of the stochastic differential
equations for position and momentum, because it allows an easy
visualization of the problem and also a straightforward connection
with the stochastic models for turbulence. It has been shown that
the transport of energy, typical of turbulence, can only be
obtained if a damping term is considered in the stochastic
equations (and equivalently in the master equation - ME)
\cite{Novikov}. The energy gained in the short length scale is
actually transferred to smaller ones until it is finally
dissipated.

A master equation with a dissipation term is characteristic of the
newer versions of the decoherence model \cite{Zurek}, which are
usually associated with Brownian motion \cite{Caldeira}. We show
that its correspondent stochastic differential equations for
position and momentum describe a Brownian motion only for long
time, but turbulent diffusion is actually obtained for short time.

\section{GRW model and Decoherence model}

In the GRW/GPR model the master equation is now written as

\begin{equation}
i \hbar \frac{\partial \rho (x,y,t)}{\partial t} =\left[
-\frac{\hbar ^2 }{2 m} \left( \frac{\partial ^2 }{\partial x^2}
-\frac{\partial ^2 }{\partial y^2} \right)  -i \hbar \lambda (1 -
e^{-(\alpha /4) (x - y )^2 })\right] \rho (x,y) . \label{GRW}
\end{equation}
The localization scale $1/\sqrt{\alpha } $ and the frequency of
collapse $\lambda $ are parameters chosen in such a way that the
new evolution equation does not give different results from the
usual unitary evolution for microscopic systems with few degrees
of freedom, but when a macroscopic system is described there is a
fast decay of the macroscopic linear superpositions which are
quickly transformed into statistical mixtures \cite{GRW}. Apart
from some attempts to associate the new term to gravitation
\cite{Pearle}, it has been hard to explain its nature and origin
and the values of the parameters introduced are otherwise
arbitrary.

Considering a free particle, the master equation obtained for the
decoherence model in the limit of high temperature is written as

\begin{equation}
i \hbar \frac{\partial \rho (x,y,t)}{\partial t} =\left[
-\frac{\hbar ^2 }{2 m} \left( \frac{\partial ^2 }{\partial x^2}
-\frac{\partial ^2 }{\partial y^2} \right) -i\hbar \frac{\gamma
}{2} (x-y) \left( \frac{\partial  }{\partial x} -\frac{\partial
}{\partial y} \right) -i \frac{mkT\gamma }{\hbar }(x-y)^2
 \right] \rho (x,y,t) ,
\label{dec}
\end{equation}
where $\hbar $ is Planck's constant, $m$ is the particle mass,
$\gamma ^{-1}$ is the relaxation time, $k$ is Boltzmann's constant
and $T$ is the temperature of the heat bath.

Contrary to eq.(\ref{GRW}) this master equation has a dissipation
term, though it is also subjected to some criticisms
\cite{crit,Gallis1}.

\subsection{Stochastic differential equations for the GRW model}

Describing a system with stochastic equations is very attractive,
because of their clarity and simplicity. In order to obtain the
corresponding stochastic differential equations for the GRW model,
we first derive the Fokker-Planck equation (FPE) in phase space by
using the Wigner transform of the density matrix

\begin{equation}
W_E (x,p,t) = \frac{1}{2\pi \hbar } \int _{-\infty}^{\infty} dq
\exp (-\frac{ipq}{\hbar })\rho (x+q/2, x-q/2,t).
\end{equation}

Assuming $(1 - e^{-\alpha (x - y )^2 /4}) \sim \alpha (x-y)^2 /4$
and $\varepsilon = \hbar ^2 \lambda \alpha /2m^2 $ we obtain

\begin{equation}
\frac{\partial W_E}{\partial t} = - \frac{p}{m} \frac{\partial
W_E}{\partial x} + \frac{\varepsilon m^2}{2 } \frac{\partial ^2
W_E}{\partial p^2} . \label{FPEno}
\end{equation}

The same FPE is obtained for a conditional probability function
$W_C (x,p,t|x_0) $, where $\int W_C (x,p,t|x_0) d x_0 =
W_E(t,p|x)$ \cite{Gardiner}. The stochastic process described by a
conditional probability satisfying the FPE is equivalent to the
following It\^o stochastic differential equations for momentum and
position

\begin{equation}
dp = \sqrt{m^2 \varepsilon } \xi (t) dt \label{dp2},
\end{equation}
and

\begin{equation}
dx = \frac{p}{m} dt \label{dx2},
\end{equation}
where the correlation function for the white noise is $<\xi (t)
\xi (t')> = \delta (t-t')$.

Since there is no damping term in the FPE, the momentum has only a
noise term which implies that

\begin{equation}
<p^2>_C = <p_{0}^2>_E + m^2 \varepsilon t \label{turb2}
\end{equation}

\begin{equation}
<x^2>_C = <x_{0}^2>_E + \frac{<p_{0}^2>_E}{m^2} t^2 +
\frac{\varepsilon }{3} t^3 .\label{turb1}
\end{equation}
The integration for position becomes simple if we apply the
identity $\int _{0}^{t} dt' \int _{0}^{t'} dt" = \int _{0}^{t} dt"
\int _{t"}^{t} dt'$ \cite{Honey}.

This cubic dependence for $<x^2>_C $, also obtained by GRW
\cite{GRW}, indicates turbulence, though they had not realized it.
This analogy has only been made recently \cite{Kusne,euro,PRA}.
Had they noticed this and the development of these models might
have followed a different direction, probably trying to justify
the origin of the added noise with a chaotic medium instead of
gravitation. However GRW did notice the violation of energy caused
by the time dependence of $<p^2>_C $, though they consider this
problem negligible since the chosen parameters lead to a very
small increase of energy. Notice that old versions of decoherence
\cite{Zeh} did not have a damping term either and even though we
cannot talk of non conservation of energy, since the system in
this case is open, its energy does increase. We will see in the
next subsection that the addition of a damping term in the
stochastic equations (and similarly in the ME) can in fact be
negligible in the short time limit of open systems, but it is
necessary for a consistent and generic model and is essential in
the energy transport typical of turbulence.

\subsection{Stochastic models for turbulence}

In a completely different field, fluid mechanics, there were
attempts to construct stochastic models of turbulence that
departed from equations similar to the differential equations
(\ref{dp2}) and (\ref{dx2}). Differently from the Brownian motion,
which has a linear dependence on time for $<x^2>$, turbulence has
a cubic dependence, but a linear dependence on time reappears for
$<p^2>$. Because of this, the first attempts to build stochastic
models for turbulence tried to see it as a Brownian diffusion in
the momentum space \cite{Yaglom} and therefore just considered a
noise term in the stochastic equation for momentum as in
eq.(\ref{dp2}). These models failed to describe the main feature
of turbulence, which is the energy transport between the motions
of various scales \cite{Novikov,Novikov94}. The only way to
recover this transport, and therefore to not contradict the
Navier-Stokes equation (the fundamental equation of turbulence,
equivalent in importance to the Schr\"odinger equation in quantum
mechanics), is by considering a damping term in the FPE or
equivalently, by adding a momentum dependent term in its
stochastic equation. We should then work with

\begin{equation}
dp = a(p)dt + \sqrt{m^2\varepsilon } \xi (t) dt \label{dpdec},
\end{equation}

\begin{equation}
dx = \frac{p}{m} dt \label{dxdec},
\end{equation}
where $a(p)$ is a function of momentum chosen as $-\gamma p$. The
justification for this choice will be clear soon when we discuss
the energy transport.

These equations lead to the following new FPE

\begin{equation}
\frac{\partial W_C}{\partial t} = - \frac{p}{m} \frac{\partial
W_C}{\partial x} + \gamma \frac{\partial (p W_C)}{\partial p} +
\frac{m^2 \varepsilon }{2} \frac{\partial ^2 W_C}{\partial p^2} .
\label{FPEwith}
\end{equation}
This is exactly the equation we obtain by doing the Wigner
transform to the decoherence master equation (\ref{dec}) and
assuming $\varepsilon = 2\gamma k T/m$. This FPE is usually
associated with Brownian motion, but this is so only for a large
time scale, on a short time scale we actually have turbulence.

From (\ref{dpdec}) and (\ref{dxdec}) we have

\begin{equation}
p(t) = p_0 e^{-\gamma t} +  \sqrt{m^2 \varepsilon } \int _{0}^{t}
e^{-\gamma (t-t')} \xi (t') dt',
\end{equation}

\begin{equation}
x(t) = x_0 + \frac{1}{m} \int _{0}^{t} p(t') dt',
\end{equation}
which lead to the quadratic averages

\begin{equation}
<p^2(t)>_C = <p_0 ^2>_E e^{-2 \gamma t} + \frac{m^2 \varepsilon
}{2\gamma } (1- e^{-2 \gamma t}), \label{p2}
\end{equation}

\begin{equation}
<x^2(t)>_C = <x_0 ^2>_E + \frac{<p_0 ^2>_E}{m^2 \gamma ^2} (1-
e^{- \gamma t})^2 + \frac{\varepsilon }{2 \gamma ^3} \left[
2\gamma t -3 +4 e^{-\gamma t} - e^{-2\gamma t}\right] .
\end{equation}

On a large time scale the system attains equilibrium, $<p^2(t)>_C
\rightarrow m^2 \varepsilon/(2\gamma )$, showing that energy is
conserved, and $<x^2(t)>_C \rightarrow \varepsilon t/\gamma ^2$,
which is the typical result for the Brownian motion.

On a short time scale we recover the time dependence typical of
turbulence

\begin{equation}
<p^2(t)>_C \rightarrow <p_0 ^2>_E + m^2 \varepsilon t ,
\end{equation}

\begin{equation}
<x^2(t)>_C \rightarrow <x_0 ^2>_E + \frac{<p_0 ^2>_E}{m^2} t^2 +
\frac{\varepsilon }{3} t^3 .
\end{equation}

Notice that from the derivative of eq.(\ref{p2}) we can calculate
the critical damping value. If $\gamma > m^2 \varepsilon /(2<p_0
^2>_E)$ the system loses energy and when $\gamma < m^2 \varepsilon
/(2<p_0 ^2>_E)$ the system gains energy. On the short time scale
the absence of the parameter $\gamma $ in the equations above
confirms that in this limit the dissipation term can be neglected.
In any case, in the large time limit ($t\gg \gamma ^{-1}$), energy
is conserved, as remarked above.

The relaxation term added to the stochastic equation for momentum
is responsible for the transportation of the injected energy to a
much smaller scale, where it is dissipated. In other words, energy
always ends up being dissipated in the environment. In the
language of turbulence $\varepsilon $ is the energy injected into
the fluid per unit time and unit mass, in an intermediary length
scale - the inertial regime - it is transferred to smaller scales
and is finally dissipated in a much smaller length scale, called
Kolmogorov dissipation scale \cite{Frisch}. Considering that we
are in the inertial regime, even though the dissipation exists, we
cannot see it, but we are able to verify the transport of energy.

The moment of third order of velocity $v$ represents the energy
transport between the motions of various scales. Its relation with
$\varepsilon $ was derived by Kolmogorov from the Navier-Stokes
equation and is given by his classical expression \cite{Frisch}

\begin{equation}
<v^3>_E = - c_{1} \varepsilon x \label{Kolmo}.
\end{equation}

Multiplying the new FPE for $W_E(t,p|x)$ by $p^2$ and integrating
over $p$ we have

\begin{equation}
\frac{\partial <p^2>_E}{\partial t} + \frac{1}{m} \frac{\partial
<p^3>_E}{\partial x} - 2<p a>_E = \varepsilon m^2 \label{Euler} .
\end{equation}

Using Kolmogorov's law and the fact that the first term of
(\ref{Euler}) is zero for statistically stationary turbulence
\cite{Novikov} (notice that $<p^2>_E $ does not depend on time in
statistically stationary turbulence, but $<p^2>_C $ does), we
arrive at

\begin{equation}
-c_{1} m^2 \varepsilon - 2 <pa>_E = m^2 \varepsilon ,
\end{equation}
which can only be satisfied if $a$ depends on $p$ with a negative
coefficient, such as $a = -\gamma p$ and $<p^2>_E = c_{2} \gamma
^{-1} m^2 \varepsilon $, with $2 c_{2} - c_{1} = 1$. We can see
that Kolmogorov's law justifies the added relaxation term in
eq.(\ref{dpdec}).

It is now clear that the FPE (\ref{FPEwith}) describes turbulence
and Brownian motion. It is the time scale that determines which
kind of diffusion we actually have. When dealing with quantum open
systems and the transition quantum/classic we should keep in mind
two time scales: the decoherence time and the relaxation time and
any proposed experiment should take them into account.

\section{Conclusions}

The connection between these two different fields is very rich and
there is certainly a lot more to be done. They can give important
contributions to each other. From the study of turbulence we have
shown that the energy gained or lost by an open system is always
dissipated in the environment. It should also be mentioned that
eq.(\ref{dec}) can also be obtained by a more generic method
developed by Kusnezov, Bulgac and Do Dang (KBD) \cite{Kusne},
where they try to give a better justification to the origin of
irreversibility in the master equation. Both system and
environment have a dynamics and the system evolves in a chaotic
background, whose dynamic evolution is described by the random
matrix theory. This may indicate that the technique of random
matrices can be useful in the study of turbulence, a field whose
main equation has always been the Navier-Stokes equation, a purely
phenomenological equation. Departing from the master equation
obtained with the KBD method, we can interpret the corresponding
stochastic differential equations as describing a tracer immersed
in a chaotic medium. We intend to extend this analogy between
turbulence and the KBD method in a future publication, where we
also intend to look for the stochastic equations associated with
the fractional Fokker-Planck equation obtained with this method
\cite{Kusne1} .

\acknowledgments The authors acknowledge the support of Funda\c
c\~ao de Amparo \`a Pesquisa do Estado de S\~ao Paulo, FAPESP.

\end{document}